# An Area-Efficient FPGA Overlay using DSP Block based Time-multiplexed Functional Units

Xiangwei Li*, Abhishek Kumar Jain*, Douglas L. Maskell* and Suhaib A. Fahmy†
*School of Computer Engineering, Nanyang Technological University, Singapore
*{xli045, abhishek013}@e.ntu.edu.sg, asdouglas@ntu.edu.sg
†School of Engineering, University of Warwick, United Kingdom
†s.fahmy@warwick.ac.uk

*Abstract*—Coarse grained overlay architectures improve FPGA design productivity by providing fast compilation and software-like programmability. Throughput oriented spatially configurable overlays typically suffer from area overheads due to the requirement of one functional unit for each compute kernel operation. Hence, these overlays have often been of limited size, supporting only relatively small compute kernels while consuming considerable FPGA resources. This paper examines the possibility of sharing the functional units among kernel operations for reducing area overheads. We propose a linear interconnected array of time-multiplexed FUs as an overlay architecture with reduced instruction storage and interconnect resource requirements, which uses a fully-pipelined, architecture-aware FU design supporting a fast context switching time. The results presented show a reduction of up to 85% in FPGA resource requirements compared to existing throughput oriented overlay architectures, with an operating frequency which approaches the theoretical limit for the FPGA device.

## I. INTRODUCTION

Coarse grained FPGA overlay architectures have emerged as an attractive solution for improving design productivity by offering fast compilation and software-like programmability. Recently, coarse grained overlay architectures have been paired with general purpose processors as a co-processor [16], [3] as this allows the hardware fabric to be viewed as a software-managed hardware task, enabling more shared use. Other advantages include application portability across devices, better design reuse, and rapid reconfiguration that is orders of magnitude faster than partial reconfiguration on fine-grained FPGAs. An overlay overcomes the need for a full cycle through vendor-implementation tools, instead presenting a simpler problem of programming an interconnected array of functional units (FUs). Overlays are not intended to replace HLS and vendor-implementation tools and are instead intended to support FPGA usage models where programmability, abstraction, resource sharing, fast compilation and design productivity are critical issues.

A number of coarse grained overlay architectures with similarities to coarse-grained reconfigurable architectures (CGRAs) have been proposed for FPGAs. These overlays and CGRAs can be categorized based on their interconnect configuration using the classification in [17], where 4 different categories are defined as: spatially configured, time multiplexed, packet switched, and circuit switched. While examples of packet switched [17] and circuit switched [12] networks in FPGAs exist, they are generally resource hungry, and are less suited for FPGA-based overlays. Thus, the majority of overlays are restricted to two classes: spatially configured; and time-multiplexed, where both the FU and the interconnect can fall within either of these two categories.

The largest group consists of spatially configured FUs and spatially configured interconnect networks [22], [2], [8], [14], [13], [15], which we refer to as SCFU-SCN. In an SCFU-SCN overlay, an FU executes a single arithmetic operation and data is transferred between FUs over a programmable, but temporally dedicated, point-to-point link. That is, both the FU and the interconnect are unchanged while a compute kernel is executing. This results in a fully pipelined, throughput oriented datapath executing one kernel iteration per clock cycle, thus having an initiation interval (II) [1] between kernel data packets of one. A number of different spatially configured interconnect strategies exist, with the most common being: island style [22], [14], [13], [15], NN [8] and to a lesser extent linear interconnect [11], [7].

The area overheads of SCFU-SCN overlays, in particular the large interconnect resource requirements, have limited the use of these overlays to very small compute kernels in practical FPGA-based systems [2]. This means that in a typical application a number of different kernels would need to be mapped to the overlay as the application executes to achieve the best application acceleration. Thus the kernel context switch time is also an important consideration in the efficient operation of an overlay [11]. Some of the current overlays utilize partial reconfiguration to reduce the overlay area, in particular the interconnect resources, by trading off runtime connection flexibility [20]. However, while faster than a complete FPGA reconfiguration, partial reconfiguration still results in a significant context switch overhead, which significantly impacts an application's runtime.

A number of overlays have been proposed which share the functional units among kernel operations in an attempt to reduce overlay resource requirements. Time-multiplexing the FU can significantly reduce the FU and interconnect resource requirements but at the cost of a higher II and hence a reduced throughput. The most common time-multiplexed overlays have both a time-multiplexed FU and a time-multiplexed interconnect network [19], [21], [5], [20], which we refer to as TMFU-TMN. Time-multiplexing the overlay allows it to change its behavior on a cycle by cycle basis while the compute kernel is executing [19], [21], [5], [20], thus allowing sharing of limited FPGA resources. However, in many cases the storage requirements for instructions are very large, resulting in a significant area overhead. This is mainly due to: the scheduling strategy; the execution model; and the design of overlay architecture, and limits the scalability of the overlay while also impacting the kernel context switch time. Additionally, many of these overlays are not architecture-focused and hence the FU operates at a relatively slow frequency.

The design of an efficient overlay should address the concerns identified above. Thus, in this paper, we present a TMFU-TMN overlay with reduced area overheads due to a reduction in the instruction storage and interconnect resource requirements. The main contributions are:

- An architecture using a linear connection of FUs to form a processing pipeline which is able to support the execution of feed-forward data flow graphs (DFGs). This architecture results in a significant reduction in both the FU and interconnect compared to other overlays.
- A scheduling methodology which reduces the number of instructions that need to be stored in each FU thus enabling an FU with a very small memory footprint.
- A 32 bit area-efficient, time-multiplexed FU with a very small instruction memory built around a fully pipelined DSP block using just a few memory primitives.
- An architecture with a significantly reduced context switching time, thus enabling rapid kernel changes.

This combination of a small, highly pipelined FU with a simple linear interconnect allows us to achieve a high operating frequency and a relatively fast kernel context switch time but at the expense of an increased II.

## II. Related Work

In TMFU CGRA overlays there are many options for the type of FU, however, in FPGAs, the choice is more limited. Here an FU could be a soft RISC processor [6], a DSP macro based soft processor [9] or some simple ALU like structure [19]. The FU would typically have an instruction memory, allowing multiple operations to be time-multiplexed to the FU. As with SCN interconnect networks, there are many possibilities, with the NN style of programmable interconnect, being most common. One benefit of using these types of overlays is that well established algorithms and tools can be used for application mapping.

One example is CARBON [5], which was implemented as a $2\times2$ array of tiles on an Altera Stratix III FPGA. Each tile has an FU with a programmable ALU and instruction memory, supporting up to 256 instructions. An FU consumed 3K ALMs, 304 FFs, 15.6K BRAM bits and 4 DSP blocks, achieving an operating frequency of 90 MHz. Compared to the other TMFU-TMN overlays discussed here, CARBON has a large FU resource requirement with a relatively slow speed which limits the scalability of the architecture.

The SCGRA overlay [19] was proposed to address the FPGA design productivity issue, demonstrating a 10-100$\times$ reduction in compilation time compared to the AutoESL HLS tool. Application specific SCGRA overlays were subsequently implemented on the Xilinx Zynq platform [18], achieving a speedup of up to 9$\times$ compared to the same application running on the Zynq ARM processor. The FU used in the Zynq based SCGRA overlay operates at 250 MHz and consists of an ALU, multiport data memory (256$\times$32 bits) and a customizable depth instruction ROM (Supporting 72-bit wide instructions) which results in the excessive utilization of BRAMs. As the full FPGA bitstream needs to be reconfigured for a compute kernel change, very fast context switching between applications, in the order of a few microseconds, is not possible.

The reMORPH [20] overlay better targeted the FPGA fabric, with an FU consuming 1 DSP Block, 3 block RAMs, 196 LUTs and 41 registers. To reduce the overhead due to routing and muxes, the reMORPH FU does not use decoders resulting in a 72-bit instruction memory (supporting up to 512 instructions) which causes an over utilization of BRAMs. Tiles are interconnected using an NN style of non-programmable interconnect, which is adapted using partial reconfiguration at runtime, and hence, suffers from the same slow hardware context switch problem as SCGRA.

The TILT overlay [21], being a floating point overlay, unlike the other overlays discussed here, results in high resource consumption. An 8-core TILT system (with each core having one multiply FU and one add/sub FU) was specifically designed to implement a 64-tap FIR filter application, resulting in a throughput of 30 M inputs/sec and consuming 12K eALMs. For the same application, Altera OpenCL HLS was used to generate a fully parallel and pipelined implementation, resulting in a throughput of 240 M inputs/sec (8$\times$ higher throughput than 8-core TILT) and consuming 51K eALMs (4$\times$ higher area than 8-core TILT).

Most of the time multiplexed overlays described above suffer from large area overheads due to instruction storage requirements, which along with their use of partial reconfiguration results in a long kernel context switch time. Thus, there is significant scope to develop a time multiplexed overlay which is able to better address these issues.

## III. Programmable Pipelines

A fully pipelined SCFU-SCN overlay can deliver maximum performance by executing one computation iteration every clock cycle (that is it has an II of one), but with a large FPGA resource overhead. Alternatively, a TMFU-TMN overlay with its reduced FPGA resource requirements may be a feasible alternative allowing the remainder of the FPGA fabric to be utilized for other purposes. However, the exact architecture needs to be carefully analysed taking into account the characteristics of the application kernels and the underlying FPGA architecture, so as to minimise FPGA resource usage.

As discussed earlier, some of the more area efficient overlays utilise a simple linear interconnect structure, which can reduce to a simple direct connection between FUs by allocating DFG nodes from the same scheduling time step to the individual FUs. For example, Fig. 1(a) shows the medical imaging 'gradient' benchmark [10], while Fig. 1(b) shows the resulting data flow graph (DFG). By using a simple ASAP schedule, we can allocate the nodes in each stage to a different FU which in this example results in 4 stages, with the FU in each stage being time-multiplexed among stage operations using a direct (non-programmable) connection between FUs. That is, the first stage would contain four subtract operations which would execute on the first FU, the second stage would contain four multiplication operations executing on the second FU, and so on. Thus for the example shown in Fig. 1(b) the II would be 11, consisting of 5 cycles for data entry, 4 cycles for the 4 subtract operations, 1 cycle for data output and 1 cycle to flush the pipeline. Note that multiplexing the kernel operations of the DFG in Fig. 1(b) to a single FU would result in an II of 17 (5 load, 11 operation, and 1 store), assuming best case execution without NOP insertions, while a spatially configured overlay would require 11 FUs with an II of 1.

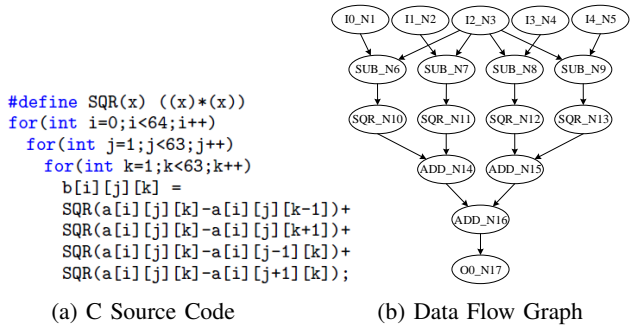

(a) C Source Code  (b) Data Flow Graph

Fig. 1: The 'gradient' benchmark.

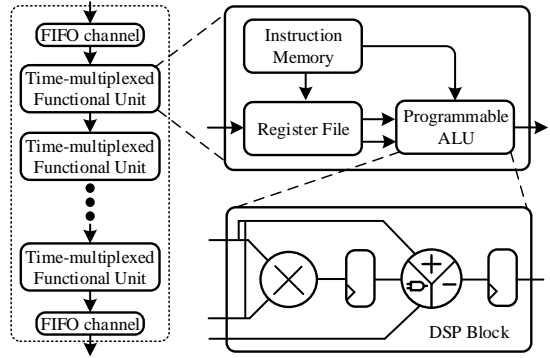

Fig. 2: A programmable processing pipeline.

Furthermore, this means that only a small subset of all possible instructions needs to be stored in each FU, resulting in a very small memory requirement overcoming the problem of the large area overheads due to the instruction storage requirements in the overlays described in Section 2. This is further assisted by using a simple low resource, but highly pipelined, FU (similar to the iDEA soft core processor [9]). This combination of linear interconnect and time multiplexed FU presents an interesting design space which will likely result in high throughput programmable architectures with reduced II and significantly reduced hardware resource requirements.

*A. Architecture Description*

The 32-bit pipeline consists of a streaming data interface made up of Distributed RAM (DRAM) acting as a FIFO, which feeds into a cascade of time-multiplexed FUs, with another DRAM-based FIFO at the pipeline output, as shown in Fig. 2. The FU is optimized for FPGA implementation, and consists of a lightweight processor, based loosely on the iDEA DSP48E1-based soft core processor [9]. The main FU components are an instruction memory (IM), register file (RF) and DSP-based ALU, as shown in Fig. 3.

At initialization, or upon a hardware context switch, a 40-bit data word, made up of a 32-bit wide instruction and an 8-bit tag (used to match an instruction with its corresponding FU), is clocked to the FU instruction port from a separate 40-bit wide context memory implemented externally using BRAMs.

The FU instruction ports are daisy-chained together, allowing for efficient configuration of the complete overlay. There are two types of instruction: arithmetic or data bypass (used to forward data to the next stage). A 5-bit instruction counter (IC) is used to keep track of the instructions written to the FU. There are two types of instruction: arithmetic or data bypass. The FU architecture supports a 32 entry IM implemented using RAM32M primitives. The RAM32M primitive is an efficient memory primitive implemented in LUTRAMs, which can be configured as a 32 deep 2-bit wide quad port (3 read, 1 read/write), 32 deep 4-bit wide dual port (1 read, 1 read/write) or an 8-bit wide single port (1 read/write) memory. Since the IM writes only occur once at context initialization, we multiplex the read and write addresses enabling us to use the single port variant which uses just four RAM32M primitives to instantiate the $32{\times}32$ IM. A 32-bit instruction has two parts, the 21-bit DSP block configuration and two 5-bit source operand addresses. Upon the completion of the context write cycle, each FU contains the necessary instructions in its IM and the instruction count in its IC register.

The RF is used to hold the 32-bit data streamed from the input FIFO or the previous FU stage. The RF requires two read ports and one write port, but as reads and writes do not occur simultaneously we again multiplex one port enabling both read and write operations. Thus the 32 entry RF requires 8 RAM32M primitives configured as a 1 read/write and 1 read port memory. Data is streamed into the FU when the *valid* signal is high and is written to the RF using a simple sequential data counter (DC). When all the data is available in the RF the *valid* signal is taken low. The control generator then asserts a *control* signal which triggers the IM to start sending operand addresses to the RF and configuration data to the DSP block. A program counter (PC) controls the execution of instructions while the *control* signal is high. After executing all the instructions related to the scheduling stage, including the data output to the next stage, or output FIFO, the DSP block flushes its internal pipeline and the program counter resets allowing the same sequence of instructions to be reissued. Thus, each FU only needs to execute a small number of instructions, allowing for a small IM and RF design.

The other major block in the FU is the ALU, which consists of a DSP48E1 primitive, two 32-bit registers, one at the C input port for pipeline balancing and another at the output, and an 18-bit register for holding the ALU configuration data. The DSP48E1 primitive can support various operations determined by the configuration control inputs. As instruction decoders are not used the instruction format must explicitly specify the read addresses and the modes of operation of the DSP block directly, allowing us to achieve a relatively high frequency.

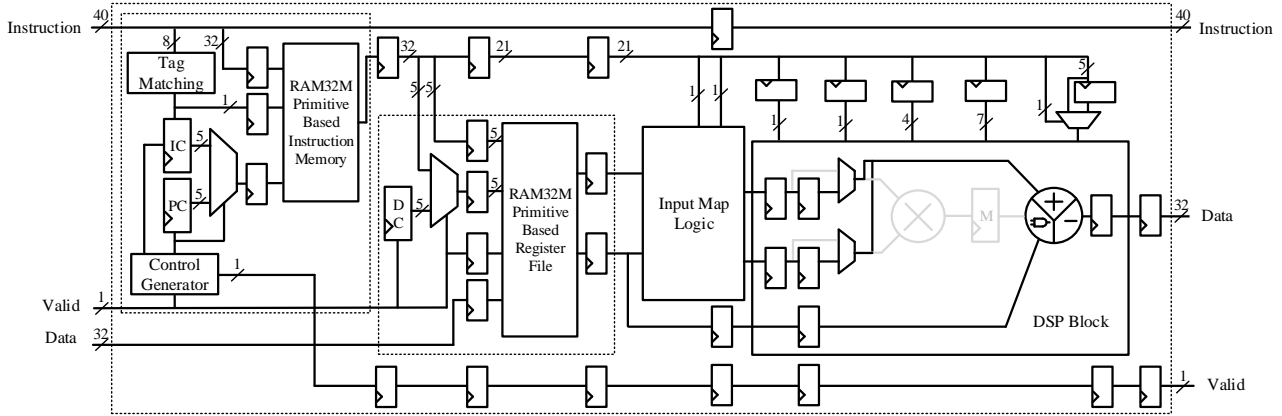

Fig. 3: The proposed time-multiplexed functional unit.

The FU of Fig. 3 was synthesized and mapped to a Xilinx Zynq XC7Z020-1CLG484C using Xilinx ISE 14.6. We achieve a frequency of 325 MHz while consuming 1 DSP block, 160 LUTs and 293 FFs. A complete pipeline consisting of 8 FUs and the 2 I/O FIFOs, consumed 8 DSP blocks, 808 LUTs and 1077 FFs representing less than 4% of the Zynq FPGA resources while operating at a slightly reduced frequency of 303 MHz. When mapped to a more capable XC7VX485T Virtex 7 device, we achieve a frequency in excess of 600 MHz for the same resource utilization. The maximum configuration time for a single pipeline consisting of 8 FUs is 0.85 µs at 300 MHz. This assumes that all 8 IMs require the full 32 instructions and that the kernel contexts are already preloaded into the external context memory.

To overcome the effect of the increase in II compared to a spatially configured overlay, we propose replicating multiple pipelines, as shown in Fig. 4. This way, the proposed pipeline can be used as a high performance hardware accelerator which is integrated within a general purpose processor based system, such as the Xilinx Zynq SOPC. Integration within the SOPC allows the use of an OS or hypervisor for runtime management of both the software tasks on the general purpose processor and the hardware tasks on the proposed overlay using software APIs. As shown in Fig. 4, a memory subsystem is required as a bridge between the overlay on the FPGA fabric, the ARM processor and the external memory. This memory subsystem consists of a single port Block RAM for each programmable pipeline and a single Block RAM for configuration data for all pipelines. Data transfer between these memories and the external memory is performed under DMA control.

## IV. COMPILING TO THE OVERLAY

An overlay has two separate design processes: Overlay implementation on the FPGA and application mapping to the overlay. While the design and implementation of the overlay relies on the conventional hardware design flow using vendor tools, this process is done offline, once only, and so does not impact the compute kernel implementation of an application. We then use an in-house automated compilation flow to provide a rapid, vendor independent, mapping to the overlay. The mapping process comprises DFG extraction from high level compute kernels, scheduling of the DFG nodes onto the overlay, and finally, the instruction generation for each FU. This is also done offline. Then at power-on the bitstream for the overlay, and any other unrelated hardware components, is used to configure the FPGA. Subsequent to this, the ARM processor loads the kernel configuration into the overlay pipeline and initiates kernel execution. Our mapping flow is described below using the previous example.

*HLL to DFG Conversion:* The tool transforms a 'C' description of the compute kernel to a DFG text description, where nodes represent operations and edges represent data flow between operations, as shown in Fig. 1(b).

*Operation Scheduling:* Scheduling is used to generate a sequenced DFG, with nodes in each scheduling stage being allocated to a single FU for execution. Here, the set of instructions from the sequenced DFG is identified, then the cycle-by-cycle execution pattern is formed as shown in Table. I, and lastly the 32-bit FU instructions are generated.

For the example shown in Fig. 1, FU0 waits for an initial 5 cycles (from clock cycle 1 to clock cycle 5) while the data is loaded into the RF, as indicated in Table I. As soon as loading has completed, FU0 is triggered (in the 6th clock cycle) and starts executing the 4 SUB instructions, one every cycle, using the data from the RF. Note that as all operations are in the same scheduling stage there is no data dependency between operations. FU0 starts sending the resulting data to FU1 on the 8th clock cycle (due to the 3 stage internal pipeline in the DSP block) and then waits for the cycle to repeat. Two extra cycles (clock cycle 10 and 11) are needed for flushing the pipeline of FU0, and hence we use a back-pressure signal from FU0 to the input FIFO channel (from clock cycle 6 to clock cycle 11) to pause further data input.

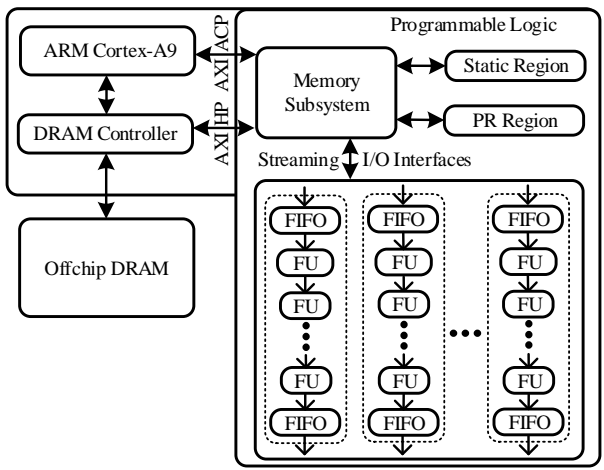

Fig. 4: The proposed overlay on the Zynq platform.

TABLE I: First 32 cycles of the schedule with II=11.

| cycle | FU0 | FU1 | FU2 | FU3 |
|---|---|---|---|---|
| 1 | Load R0 | | | |
| 2 | Load R1 | | | |
| 3 | Load R2 | | | |
| 4 | Load R3 | | | |
| 5 | Load R4 | | | |
| 6 | SUB (R0 R2) | | | |
| 7 | SUB (R1 R2) | | | |
| 8 | SUB (R2 R3) | Load R0 | | |
| 9 | SUB (R2 R4) | Load R1 | | |
| 10 | | Load R2 | | |
| 11 | | Load R3 | | |
| 12 | Load R0 | SQR (R0 R0) | | |
| 13 | Load R1 | SQR (R1 R1) | | |
| 14 | Load R2 | SQR (R2 R2) | Load R0 | |
| 15 | Load R3 | SQR (R3 R3) | Load R1 | |
| 16 | Load R4 | | Load R2 | |
| 17 | SUB (R0 R2) | | Load R3 | |
| 18 | SUB (R1 R2) | | ADD (R0 R1) | |
| 19 | SUB (R2 R3) | Load R0 | ADD (R2 R3) | |
| 20 | SUB (R2 R4) | Load R1 | | Load R0 |
| 21 | | Load R2 | | Load R1 |
| 22 | | Load R3 | | ADD (R0 R1) |
| 23 | Load R0 | SQR (R0 R0) | | |
| 24 | Load R1 | SQR (R1 R1) | | |
| 25 | Load R2 | SQR (R2 R2) | Load R0 | |
| 26 | Load R3 | SQR (R3 R3) | Load R1 | |
| 27 | Load R4 | | Load R2 | |
| 28 | SUB (R0 R2) | | Load R3 | |
| 29 | SUB (R1 R2) | | ADD (R0 R1) | |
| 30 | SUB (R2 R3) | Load R0 | ADD (R2 R3) | |
| 31 | SUB (R2 R4) | Load R1 | | |
| 32 | | Load R2 | | Load R0 |

TABLE II: DFG characteristics of benchmark set.

| No. | Benchmark Name | i/o nodes | graph edges | op nodes | graph depth | average parallelism | II | eOPC |
|---|---|---|---|---|---|---|---|---|
| 1. | chebyshev | 1/1 | 12 | 7 | 7 | 1.00 | 6 | 1.2 |
| 2. | sgfilter | 2/1 | 27 | 18 | 9 | 2.00 | 10 | 1.8 |
| 3. | mibench | 3/1 | 22 | 13 | 6 | 2.16 | 11 | 1.2 |
| 4. | qspline | 7/1 | 50 | 26 | 8 | 3.25 | 18 | 1.4 |
| 5. | poly5 | 3/1 | 43 | 27 | 9 | 3.00 | 14 | 1.9 |
| 6. | poly6 | 3/1 | 72 | 44 | 11 | 4.00 | 17 | 2.6 |
| 7. | poly7 | 3/1 | 62 | 39 | 13 | 3.00 | 17 | 2.3 |
| 8. | poly8 | 3/1 | 51 | 32 | 11 | 2.90 | 15 | 2.1 |

The operation of FU1 is similar to that of FU0. The output data from FU0 (the SUB instruction at cycle 6) is input to FU1 at cycle 8 due to the pipeline depth. Data input into the RF of FU1 takes 4 cycles (clock cycle 8 to clock cycle 11), before FU1 is triggered (on the 12th clock cycle) and starts executing the 4 multiply (SQR) instructions on the data available in its RF. Once the first instruction is completed, it outputs the processed data to FU2 (for 4 cycles starting from the 14th clock cycle) and waits for the next set of data from FU0, upon which it repeats the previous operation. FU2 and FU3 also execute in a similar manner.

The proposed scheduling approach allows us to store just a small number of instructions for each FU, resulting in very small memories and hence an area efficient overlay. Other time multiplexed architectures, such as in [18] need large instruction memories as the FUs in those architectures need to store the full list of instructions executing every cycle, resulting in much higher resource utilization.

## V. Experimental Evaluation

We evaluate the performance of the proposed overlay and that of our compiler using a set of compute kernels extracted from compute intensive applications from the literature [13], [4], as shown in Table II. The graph depth corresponds to the number of FUs needed in the proposed overlay, while the effective operations per cycle (eOPC) is the ratio of DFG nodes (op nodes) and the II, which ranges from 1.2 to 2.6. The II is high for benchmarks with a large number of I/O nodes and high average parallelism. But, as there are no loop carried dependencies, we can replicate the processing pipeline, as in Fig. 4, to effectively achieve a lower II.

To demonstrate the benefits of the proposed overlay, we compare it to one of the more efficient SCFU-SCN overlays from the literature [13] and to RTL implementations of the same kernels using Vivado HLS 2014.2. For all 3 implementations we use the minimal number of FUs/hardware for the benchmark implementations. This is to observe the effect of FU reduction on the area requirement. Fig. 5 shows the number of FUs required for the proposed TMFU-TMN overlay compared to that of the SCFU-SCN overlay in [13] for each of the benchmarks in Table II. Here we see a significant reduction in the number of FUs required, but at the expense of an increase in the II. In the instances requiring more than 8 FUs (that is benchmarks 2, 5, and 6-8) two of the 8 FU pipelines shown in Fig. 4 are cascaded.

Because the different implementations we are attempting to compare use differing hardware resources, it is difficult to compare them directly. Instead we use a single equivalent slices (or e-Slices) metric, where we assume that 1 DSP block is equivalent to 60 slices based on the ratio of slices/DSP on the Zynq XC7Z02-1CLG484C (which is approximate 60). That is, if the FU in the proposed overlay consumes 1 DSP block and 81 slices it uses 141 e-Slices.

Then, for the benchmarks in Table II, we obtain the area in e-Slices and the throughput in Giga-operations/s (GOPS) for the 3 implementations on a Xilinx Zynq XC7Z02-1CLG484C, as shown in Table III. The FPGA hardware area consumption for the proposed overlay shows a significant reduction compared to the SCFU-SCN overlay (up to 63% FUs and 85% e-Slices), while using just 35% more resources than the Vivado implementations, as shown in Fig. 6. The time multiplexed overlay has a significant reduction in the throughput (ranging from $6\times$ to $18\times$) compared to the SCFU-SCN overlay and the Vivado implementations, due to the larger II which is acceptable in cases when a low to moderate throughput is sufficient. The important point here is that the proposed TMFU-TMN overlay has comparable area to the Vivado implementations, but with a reduced throughput, compared to the SCFU-SCN overlay which has comparable throughput, but with an increased area and so it represents another useful design alternative for overlay design space exploration. To put things into proper perspective, we also calculate the throughput per unit area for the three different implementations in MOPS/eSlices, which ranges from 0.35-0.5, 1.04-1.48, and 4.8-11.5 for the proposed overlay, the SCFU-SCN overlay and for the HLS implementations, respectively.

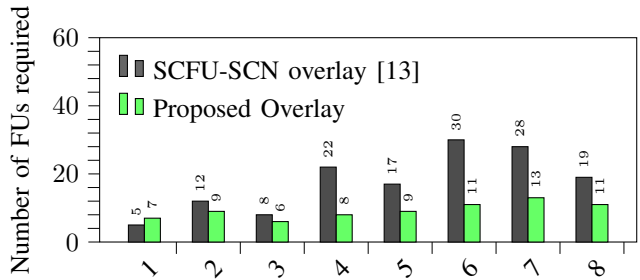

Fig. 5: Number of FUs required for the benchmarks.

TABLE III: Area and throughput comparison.

| No. | Benchmark Name | Proposed Overlay Tput | Proposed Overlay Area | SCFU-SCN overlay [13] Tput | SCFU-SCN overlay [13] Area | Vivado HLS Tput | Vivado HLS Area |
|---|---|---|---|---|---|---|---|
| 1. | chebyshev | 0.35 | 987  | 2.35  | 1900  | 2.21  | 265  |
| 2. | sgfilter  | 0.54 | 1269 | 6.03  | 4560  | 4.59  | 645  |
| 3. | mibench   | 0.35 | 846  | 4.36  | 3040  | 3.51  | 305  |
| 4. | qspline   | 0.43 | 1128 | 8.71  | 8360  | 6.11  | 1270 |
| 5. | poly5     | 0.58 | 1269 | 9.05  | 6460  | 7.02  | 765  |
| 6. | poly6     | 0.78 | 1551 | 14.74 | 11400 | 11.88 | 1455 |
| 7. | poly7     | 0.69 | 1833 | 13.07 | 10640 | 10.92 | 1025 |
| 8. | poly8     | 0.64 | 1551 | 10.72 | 7220  | 8.32  | 1025 |

The context configuration data of the benchmark set for the proposed overlay ranges from 65 Bytes to 410 Bytes. Thus, the worst case context switch between kernels takes 82 cycles (using a 40-bit wide context data word), which at 300 MHz is 0.27 us. For the SCFU-SCN overlay [13] with a worst case of 323 Bytes of configuration data which takes 13 us. This is more than that of the proposed overlay because it does not use a local context memory, and instead configuration data must come from external memory, which is slower. For the HLS implementation, we assume a PR bitstream size of 75 kByte, which is just large enough for the largest benchmark. On the Zynq platform this region requires a configuration time of 200 us. Thus the proposed overlay allows for very fast context switching between kernels.

## VI. CONCLUSION

We have presented an area efficient FPGA overlay that uses a linear connection of time-multiplexed FUs based on the Xilinx DSP48E1 macro. While the overlay exhibits a lower throughput than spatially configured overlays due to the much larger II, it has a significantly reduced FPGA resource requirement and a much lower context switching time. Our experimental evaluation shows that for a range of benchmarks, the proposed overlay delivers a throughput which is $6\times$ to $18\times$ less than the SCFU-SCN and Vivado implementations, but which uses 85% fewer e-Slices than the SCFU-SCN overlay and just 35% more e-Slices than the Vivado implementations. We are currently examining architectural modifications to reduce the II, which would result in an improved throughput.

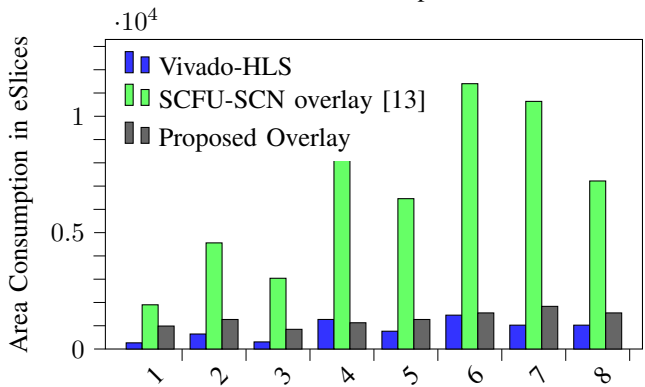

Fig. 6: Area comparison for the benchmarks.